\documentclass{aa}

\usepackage[tight]{subfigure}
\usepackage{graphicx}
\usepackage{txfonts}
\usepackage{color}

\usepackage{chemarrow}
\usepackage{multicol}

\usepackage[colorlinks]{hyperref}
\hypersetup{ colorlinks, linkcolor=blue, citecolor=blue }

\usepackage{natbib}
\bibpunct{(}{)}{;}{a}{}{,}

\begin{document}
  \title{Constraints on single-degenerate Chandrasekhar mass progenitors of Type I\lowercase{ax} supernovae}


   \author{Zheng-Wei Liu
          \inst{1},
         Takashi J. Moriya 
          \inst{1},
          Richard J. Stancliffe
          \inst{1},
            \and
           B. Wang
          \inst{2,3}
           }

   \institute{Argelander-Institut f\"ur Astronomie, Universit\"at Bonn, Auf dem H\"ugel 71, 53121 Bonn, Germany\\
            \email{zwliu@ynao.ac.cn}
         \and
              Yunnan Observatories, Chinese Academy of Sciences, Kunming 650011, P.R. China\
         \and
              Key Laboratory for the Structure and Evolution of Celestial Objects, Chinese Academy of Sciences, Kunming 650011, P.R. China\
             }

 \abstract
   {Type Iax supernovae (SNe Iax) are proposed as one new sub-class of SNe Ia since they present
    sufficiently distinct observational properties from the bulk of SNe Ia. Observationally, SNe Iax 
    have been estimated to account for $\sim$\,$5\%$--$30\%$ of the total SN Ia rate, and most SNe Iax have been discovered in 
     late-type galaxies. In addition, observations constrain the progenitor systems of some SN Iax progenitors that have ages of $<$\,$80\,\rm{Myr}$.
     Although the identity of 
     the progenitors of SNe Iax is unclear, the weak deflagration explosions of Chandrasekhar-mass (Ch-mass) 
    carbon/oxygen white dwarfs (C/O WDs) seem to provide a viable physical scenario.}
   {Comparing theoretical predictions from binary population synthesis (BPS) calculations with observations of SNe Iax, we put constraints on the 
    single-degenerate (SD) Ch-mass model as a possible SN Iax progenitor.}
   {Based on the SD Ch-mass model, the SN rates and delay times are predicted by combining binary evolution 
    calculations for the progenitor systems into a BPS model. Moreover, with current X-ray observations 
    of SNe Iax, we constrain the pre-explosion mass-loss rates of stellar progenitor systems by using two analytic models.}
   { 
     From our calculations, the long delay times of $\gtrsim3\,\rm{Gyr}$ and low SN rates of $\sim$$3\times10^{-5}\,\rm{yr^{-1}}$ are found in the red-gaint donor
     channel, indicating that this channel is unlikely to produce SNe Iax.  With our standard models, we predict that the Galactic SN Iax rate from the main-sequence (helium-star) donor 
     scenario is $\sim$$1.5\times10^{-3}\,\rm{yr^{-1}}$ ($\sim$$3\times10^{-4}\,\rm{yr^{-1}}$). The total rate of these two models is consistent with 
     the observed SN Iax rate. The short delay times in the helium-star donor channel ($<$\,$100\,\rm{Myr}$) support the young
     host environments of SNe Iax. However, the relatively long delay times in the main-sequence donor channel ($\sim$$250\,\rm{Myr}$--$1\,\rm{Gyr}$)  are less favourable for
     the observational constraints on the ages of SN Iax progenitors. Finally, with current X-ray observations for SNe Iax, we set an upper limit 
     on the pre-SN mass-loss rate at $\dot{M}\lesssim \rm{a\ few}\times10^{-4}\,\rm{M_{\odot}\,yr^{-1}}$ (for
     a wind velocity of $v_{w}=100\,\rm{km\,s^{-1}}$).}
   { The delay times in the SD Ch-mass model do not account for a significant number of SNe 
     Iax being located in late-type, star-forming galaxies. However, at least one SN Iax event (SN 2008ge) is hosted 
     by an S0 galaxy with 
     no signs of star formation. Current X-ray observations for SNe Iax cannot rule out the 
     SD Ch-mass model. Taking all these into account and considering the uncertainty of the observed rate for SN Iax events, we suggest 
     that some SNe Iax may be produced from weak deflagrations of Ch-mass C/O WDs in SD progenitor systems, especially in the helium-star donor channel.
     This is consistent with recent analysis of HST observations, which suggests
that the SN Iax SN2012Z had a progenitor system which contained a helium-star companion.  However, this 
     SD deflagration model is still unlikely to be the most common progenitor scenario for SNe Iax.  }


   \authorrunning{Z. W. Liu et al.}
   
   \titlerunning{Constraints on SD Ch-mass progenitors of SNe I\lowercase{ax}}

   \maketitle
%

\section{INTRODUCTION}
 \label{sec:introduction}

Type Ia supernovae (SNe Ia) are thermonuclear explosion of carbon-oxygen white dwarfs (C/O WDs) in binary systems.
However, the progenitor systems of SNe Ia have not yet been confidently identified and the physics of 
the explosion mechanism are still unclear (for reviews, see \citealt{Hill00, Wang12, Maoz13}). 
The spectra of most SNe Ia are characterized by strong silicon absorption features as well as the 
absence of hydrogen (H) and helium (He) lines. In addition, most SNe Ia have strong lines from intermediate-mass 
elements (IMEs) in their near-maximum-light spectra \citep{Fili97}. Although the majority of observed SNe 
Ia (normal SNe Ia) display very similar photometric and spectroscopic features, a number of events 
have several observational characteristics that are significantly distinct from the others \citep{Li01, Li11, Fole13}.

Recently, a new subclass of SNe Ia (also known as ``Type Iax supernovae'' or 
SNe Iax, see \citealt{Fole13}) has been proposed. The properties of SNe Iax are characterized by its prototypical 
member, SN 2002cx \citep{Li03}. SNe Iax have spectral and photometric properties 
that are similar to those of SNe Ia. No evidence of hydrogen is shown in their spectra. 
The ejecta of SNe Iax is dominated by IMEs and iron-group elements, which also suggests
a physical connection to normal SNe Ia \citep{Fole13}. 
All SNe Iax show clear signs of C/O-burning in their maximum-light spectra as in normal SNe Ia, 
which supports the arguments that SNe Iax are from thermonuclear explosions of
C/O WDs \citep{Fole13}. Moreover, the presence of sulphur in 
the spectra of some SNe Iax was reported as evidence for thermonuclear burning in a C/O WD \citep{Fole10b}.

SNe Iax are distinguished from SNe Ia in the following ways.
SNe Iax are significantly fainter than normal SNe Ia (\citealt{Fole13, Fole14}).  They have a wide range of explosion 
energies ($10^{49}$--$10^{51}\,\rm{erg}$), ejecta masses ($0.15$--$0.5\,\rm{M_{\odot}}$), and $^{56}\rm{Ni}$
masses ($0.003$--$0.3\,\rm{M_{\odot}}$). The spectra of SNe Iax are characterized by lower expansion velocities ($2000$--$8000\,\rm{km\,s^{-1}}$) 
than those of normal SNe Ia ($\sim$\,$15000\,\rm{km\,s^{-1}}$) at similar epochs. Their SN ejecta show strong 
mixing with both IMEs and iron-group elements in all layers \citep{Jha06, Phil07, Fole13}.
This is in clear contrast to normal SNe Ia, which are characterized by strongly
layered ejecta \citep{Mazz07}. 
Instead of entering a nebular phase dominated by broad forbidden lines of iron-peak elements, the late-time 
spectra of SNe Iax are dominated by narrow permitted Fe~II \citep{Jha06}. Moreover, two SNe Iax (SN 2004cs 
and SN 2007J) were identified  with strong He lines in their spectra \citep{Fole09, Fole13}, indicating that 
there might be He in their progenitor systems. However, \citet{Whit14} suggest that these two objects are 
likely to be core-collapse SNe rather than SNe Iax. In addition, the secondary maxima seen in normal SNe Ia have not been
observed in SNe Iax.

\begin{figure*}
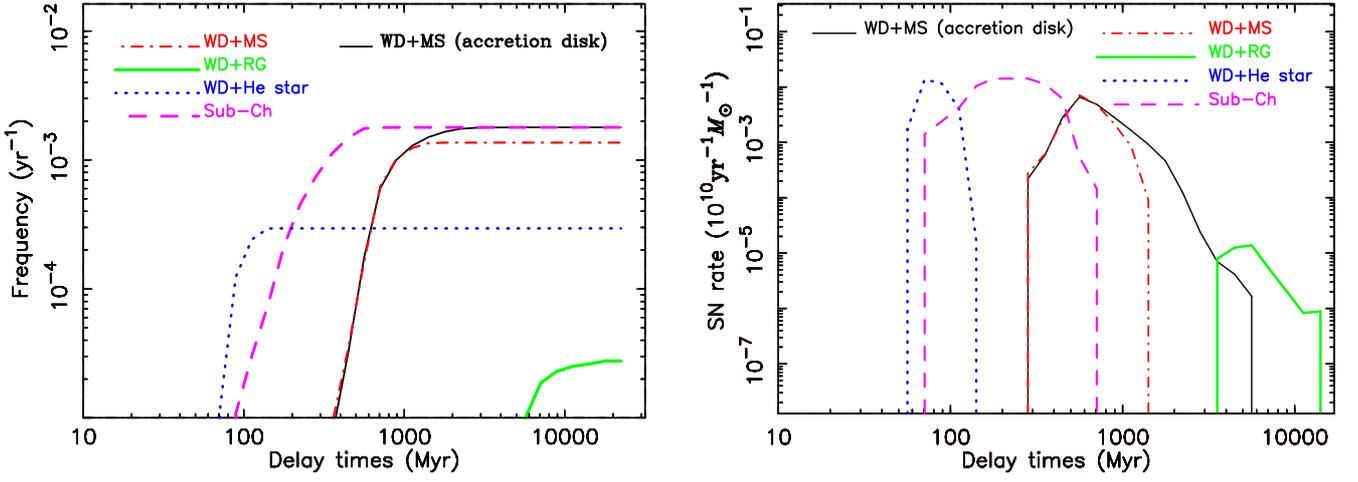

  \begin{center}
    {\includegraphics[width=0.34\textwidth, angle=270]{f1a.eps}}
\hspace{0.2in}
    {\includegraphics[width=0.34\textwidth, angle=270]{f1b.eps}}
  \caption{ Left: evolution of the Galactic SN rate of SNe Iax for a constant star 
            formation rate ($Z=0.02$, SFR=$5\,\rm{M_{\odot}\,yr^{-1}}$) in the WD+MS (dash-dotted curve), 
            WD+He star (dotted curve), and WD+RG (thick solid curve) channel. Right: as for the left, but for the single starburst case. 
            Results in the WD+MS channel with the effect of the accretion disk are shown with a thin solid line. For comparison,
            the results for the double-detonation model (i.e., the Sub-Ch model) from \citet{Wang13} are also presented with dashed lines. 
 }
\label{Fig:1}
  \end{center}
\end{figure*}

\begin{figure*}
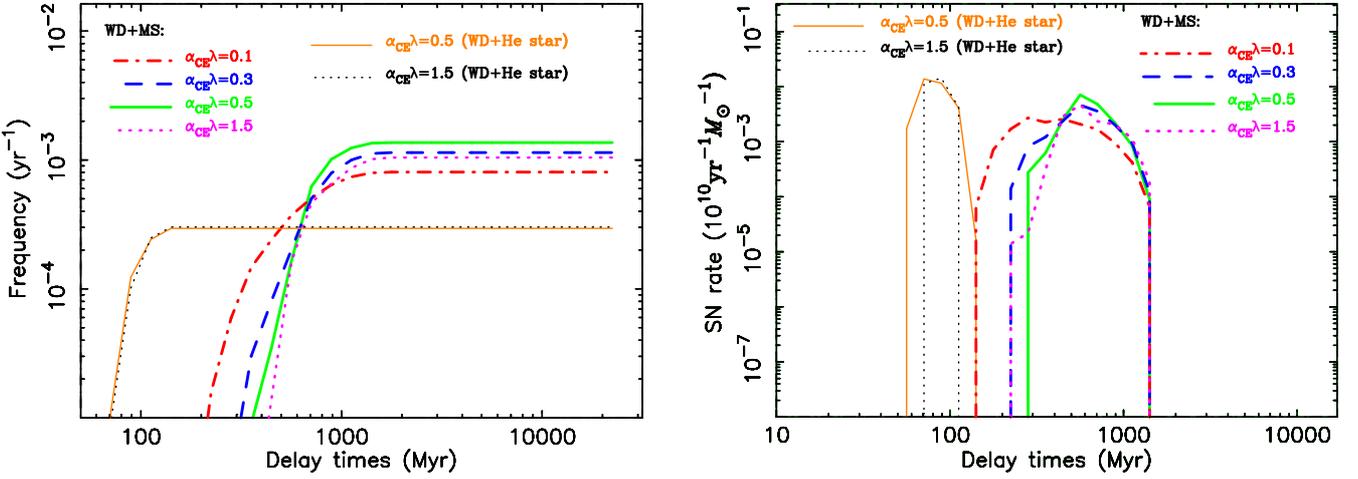

  \begin{center}
    {\includegraphics[width=0.34\textwidth, angle=270]{f2a.eps}}
 \hspace{0.2in}
   {\includegraphics[width=0.34\textwidth, angle=270]{f2b.eps}}
  \caption{ As in Fig.~\ref{Fig:1}, but only for the WD+MS (thick curves) 
            and WD+He star (thin curves) channel with different values of $\alpha_{\rm{CE}}\lambda$ (=0.1, 0.3, 0.5, 1.5).
            Only $\alpha_{\rm{CE}}\lambda$=0.5, 1.5 are studied for the WD+He star model. }
\label{Fig:2}
  \end{center}
\end{figure*}

It has been suggested that SNe Iax are the most common of all the types of peculiar 
SNe Ia (see \citealt{Fole13}). From a volume-limited sample of the Lick Observatory Supernova Search, \citet{Li11} find
that a relative rate of SNe Iax is $\sim5$ for every 100 SNe Ia. Recently, 
\citet{Fole13} have estimated that in a given volume SNe Iax could contribute $30_{-13}^{+17}\%$ of the total SN 
Ia rate. However, the more recent work of \citet{Whit14} has reduced this relative rate to  a much lower value of
$5.6_{-3.7}^{+17}\%$. \citet{Fole13} used a heterogeneous SN dataset to estimate the rate. They adopted a correction 
factor of 2 on the relative SN rate to account for various observational biases. However, \citet{Whit14} did not need to use such corrections 
because of the homogeneous nature of their survey 
and data analysis. This may be a reason for the difference in their rate estimation. A larger homogeneous sample obtained 
from future observations for SNe Iax may provide stronger constraints on the uncertainties of the SN Iax rate.

Most SNe Iax are observed in late-type, star-forming galaxies \citep{Fole13, Lyma13, Whit14}.  \citet{Lyma13} found that the host population 
of SNe Iax is similar to that of type IIP SNe, so suggest that SNe Iax have a young delay time 
of $30$--$50\,\rm{Myr}$. By studying a point source detected near the position of SN~2008ha, 
\citet{Fole14} constrained the progenitor system of SN 2008ha to have an age of $<$\,$80\,\rm{Myr}$. 
Moreover, the progenitor system of one SN Iax, SN 2012Z, has recently been detected by \citet{McCu14}. 
It was further suggested that SN 2012Z is probably a thermonuclear 
explosion of a WD accreting matter from a massive He star ($M\approx 2.0\,\rm{M_{\odot}}$ at the
time of explosion, see \citealt{McCu14}). Together with He lines seen in two SN Iax spectra, 
all the above implies relatively short delay times for the progenitor systems ($<$\,$500\,\rm{Myr}$, see also \citealt{Fole14}).  
However, there is at least one Iax event, SN 2008ge, which was observed in 
an S0 galaxy \citep{Fole10} with no signs of star formation (see \citealt{Fole13}).

Several potential progenitor systems and explosion mechanisms have been
proposed for SNe Iax \citep{Bran04, Jha06, Phil07, Fole09, Vale09, Mori10, Shen10, Fole12, Fole13, Krom13, Stri14}.
In particular, \citet{Fole13} suggest that a weak deflagration explosion of a C/O WD accreting material 
from a He-star companion may be a promising scenario for producing SNe Iax. Turbulent deflagrations in WDs 
can easily cause strong mixing of the SN ejecta \citep{Roep05}. The small amount of kinetic energy 
released from deflagration explosions is in good agreement with the 
low expansion velocities of SNe Iax.  Recently, the detailed hydrodynamics and radiative-transfer calculations for 
three-dimensional off-centre-ignited weak deflagrations of Ch-mass C/O WDs (also known 
as the ``failed deflagration model'') have showed that such deflagration explosions are able to 
reproduce the characteristic observational features of SNe Iax well (\citealt{Jord12, Krom13, Fink13}). In this
specific explosion model, a weak deflagration explosion fails to explode the entire WD, and only a part of 
the Ch-mass WD is ejected with a  much lower kinetic energy, leaving behind a polluted WD remnant  
that mainly consists of unburned C/O \citep{Jord12, Krom13, Fink13}.

In this work, we  further investigate the failed deflagration explosions of Ch-mass C/O WDs in H/He-accreting 
systems as progenitors of SNe Iax by means of binary populations synthesis (BPS) calculations. We compare 
the SN rates and delay times (times between star formation and SN explosion) from BPS calculations to 
the observations of SNe Iax. In the SD Ch-mass scenario, the circumstellar environment
is expected to be enriched by pre-SN mass loss from non-conservative mass transfer or
by the stellar wind from the donor star. After the SN explosion, X-ray emission is powered by the interaction between the SN blast wave 
and the circumstellar medium (CSM, see \citealt{Chev06, Imml06}). This X-ray emission 
can be used as a probe of the circumstellar environment and it enables us to put constraints
on the pre-SN mass-loss rates of progenitor systems \citep{Chev06, Imml06, Russ12, Marg12, Marg14}. 
We also use this X-ray diagnostics to constrain the SD Ch-mass scenario.

The theoretical SN rates and delay time distributions are predicted in Section~\ref{sec:bps}.
A parameter survey is also carried out there.  In Section~\ref{sec:csm}, we calculate 
the interaction between the SN blast wave and the CSM using simple
analytic models. The discussion of the studied progenitor scenario is given in Section~\ref{sec:discussion}. 
Finally, we summarize our results and conclude in Section~\ref{sec:summary}.

\section{Populations synthesis calculations}
\label{sec:bps}

BPS calculation has been widely used to study the evolution and 
properties of various progenitors of SNe Ia. Different SN progenitor scenarios 
involve different timescales that control the production rate of SN events, and will 
thus predict different SN age distributions. By comparing 
the theoretical Galactic birthrates and delay time distributions (DTDs) from the BPS predictions to those from
observations, we can place strong constraints on the nature of SN progenitors.

\subsection{Method}
\label{sec:code}

We investigate the SD Ch-mass explosion model, in which a C/O WD accretes material 
from a non-degenerate companion star. The companion star is potentially a main-sequence (MS), 
a slightly evolved star (WD+MS channel; \citealt{Whel73, Hach99}), a red-giant star (WD+RG channel; 
\citealt{Hach99}), or a He star (WD+He star channel, see\citealt{Wang09}). Detailed binary 
evolution calculations were performed using the stellar evolution code of \citet{Eggl71}. 
Roche-lobe overflow (RLOF) is treated by the prescription of \citet{Han00}. Additionally, 
the optically thick wind assumption of \citet{Hach96, Hach99} was used to 
describe the mass growth of a C/O WD by the accretion of H-rich material from the donor star. The 
prescription of \citet{Kato04} is implemented for the mass accumulation efficiency on to the WDs
when He-shell flashes occur. When the mass of a C/O WD got close to the Ch-mass limit ($\sim\rm{1.38\,M_{\odot}}$) 
by the accretion, we assumed that weak deflagrations are 
triggered and the C/O WD explodes as a SN Iax.  Details of how the transferred material is 
accreted onto the WD are described in Appendix~\ref{sec:appendix1}.

With a series of binary evolution calculations for various close WD binary 
systems, we determined the initial parameter space leading to SNe Iax in the orbital period--secondary 
mass (i.e., $\rm{log\,P^{i}}$--$\rm{M^{i}_{2}}$ ) plane for various WD masses in 
the WD+MS, WD+RG, and WD+He star channel. Then, all these results were incorporated into 
the BPS \citep{Hurl00, Hurl02} calculations to obtain the SN Iax 
rates and their temporal evolution for the different SD Ch-mass explosion channels. In our BPS calculations, 
the initial mass function of \citet{Mill79} is used. We assumed a circular binary orbit and
set up a constant initial mass ratio distribution (i.e., $n({q}')=$constant, see \citealt{Gold94, Bend08, Duch13}). 
We assumed that all stars are members of binaries, and the initial separation distribution used in \citet{Han95}
is adopted.\footnote{This distribution implies that the numbers of wide
binary systems per logarithmic interval are equal and that about
50 percent of stellar systems have orbital periods less than $100\,\rm{yr}$ \citep{Han95}. Recent studies indicate that
this initial separation distribution is reasonable \citep{Sana12, Duch13}.} We simply assumed a constant star formation 
rate (SFR) of $5\,\rm{M_{\odot}yr^{-1}}$ or, alternatively, a delta function, i.e., a single
starburst (see also \citealt{Wang10}). The standard energy equations of \citet{Webb84} were used to calculate the output of 
the CE phase. The CE ejection was determined with two highly uncertain 
parameters, $\alpha_{\rm{CE}}$ and $\lambda$. Here, $\alpha_{\rm{CE}}$ is the CE ejection
efficiency, i.e. the fraction of the released orbital energy used to eject the CE; $\lambda$ is a 
structure parameter that depends on the evolutionary stage of the donor star. 
In this work, the models with a metallicity of Z=0.02 and a parameter of $\alpha_{\rm{CE}}\lambda$=0.5 are defined 
to be our standard models.\footnote{Our standard models adopt the model parameters 
used in \citet{Wang09, Wang10}; however, the accretion disk of a WD is not included in our standard models 
due to its uncertainties.} If a binary system evolves to 
a WD+MS, WD+RG, or WD+He~star system and the system is located in the aforementioned orbital period--secondary 
mass plane for SNe Iax at the onset of RLOF, we then assume that a SN Iax occurs in
the system. Details on the formation of WD+MS, WD+RG, and WD+He star progenitor systems are discussed
in Appendix~\ref{sec:appendix2}. We refer to \citet{Han04} and \citet{Wang10} for detailed descriptions of our method.

\subsection{Predicted rates and DTDs of our  standard models}
\label{sec:results}

Figure~\ref{Fig:1} presents the SN Iax rates as a function of the delay times for the SD H-accreting scenario from our BPS calculations.
The SN Iax rate in the WD+MS channel is $\sim$\,$1.5\times10^{-3}\,\rm{yr^{-1}}$ ($\sim$\,$43\%$ of the total SN Ia 
rate\footnote{We assume that the total SN Ia rate is $3.5\times10^{-3}\,\rm{yr^{-1}}$  based on the inferred Galactic SN Ia rate of 
3--4$\times10^{-3}\,\rm{yr^{-1}}$ \citep{Capp97}.}), the SN Iax rate 
in the WD+RG channel is $\sim$\,$3\times10^{-5}\,\rm{yr^{-1}}$ ($\sim$\,$1\%$ of the total SN Ia rate). In addition, the SNe from the WD+MS 
channel have a range of delay times ($\sim$\,$250\,\rm{Myr}$--$1\,\rm{Gyr}$, see Fig.~\ref{Fig:1}), 
while all SNe from the WD+RG channel are quite old ($\gtrsim3\,\rm{Gyr}$).
Similarly, theoretical predictions in our standard model based on the WD+He star channel 
are also shown in Fig.~\ref{Fig:1}. The SN Iax rate in this channel is $\sim$\,$3\times10^{-4}\,\rm{yr^{-1}}$ ($\sim$\,$9\%$ of the total SN Ia rate), 
and all WD+He~star models only contribute to the SNe events with delay times shorter than $100\,\rm{Myr}$.

\begin{figure*}
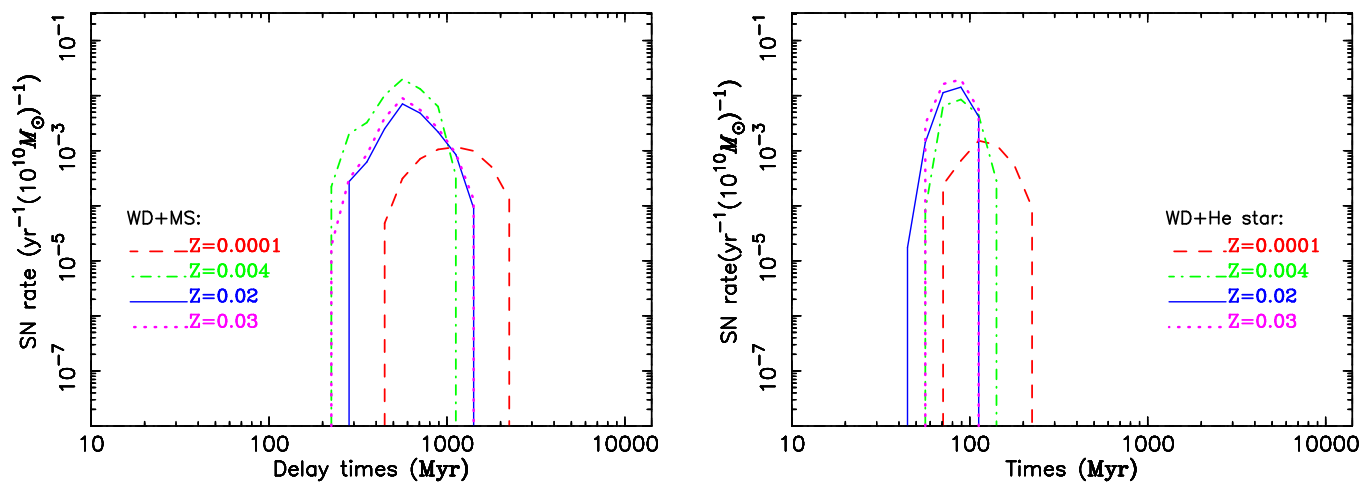

  \begin{center}
    {\includegraphics[width=0.34\textwidth, angle=270]{f3a.eps}}
 \hspace{0.2in}
   {\includegraphics[width=0.34\textwidth, angle=270]{f3b.eps}}
  \caption{Similar to the Fig.~\ref{Fig:1}, but only for the WD+MS (left panel) and WD+He 
           star (right panel) channel with different metallicities of $\rm{Z}$=(0.0001, 0.004, 0.02, 0.03).}
\label{Fig:3}
  \end{center}
\end{figure*}

\subsubsection{Comparison with the observations}

Compared with the observations, the SNe Iax from the 
WD+He~star channel contribute about $9\%$ of the total 
SN Ia rate. The delay times in the WD+He~star scenario are 
consistent with the young host environments of SN Iax progenitors. The observed 
young environments of SNe Iax are less favourable for the delay times of the H-accreting 
SD channel, especially the WD+RG channel. Considering  the very low SN rate ($\sim$\,$1\%$ of 
all SNe Ia) in the WD+RG channel, we therefore suggest that it is unlikely that SNe 
Iax are produced this way.

The SN rate from the WD+MS channel ($\sim$\,$43\%$ of the total SN Ia rate) agrees 
with the measured rate for SNe Iax ($30_{-13}^{+17}\%$ of the total SN Ia rate) reported by \citet{Fole12}, 
but much higher than the estimated rate ($5.6_{-3.7}^{+17}\%$ of the total SNe Ia rate) suggested 
by \citet{Whit14}. In addition, as mentioned above, the delay times of the WD+MS channel are difficult 
to reconcile with the observed young host environments of SNe Iax. However, some SNe Iax have been 
found in old environments \citep{Fole13}. This indicates that some SNe Iax may indeed come from the 
WD+MS channel. Therefore, it is still hard to exclude the WD+MS channel as a possible scenario 
for some SNe Iax.

By considering the uncertainties on the percentage of SNe Iax, all the above results suggest that 
some SNe Iax may come from weak deflagration explosions of Ch-mass C/O WDs in H/He-accreting SD progenitor 
systems. However, it is still difficult to ascertain that this specific deflagration explosion model is the most 
common progenitor scenario for SNe Iax because the SN rates in the SD scenario are dominated by the WD+MS 
model\footnote{A recent study of the He-accreting WDs of \citet{Pier14} suggests
that the SN rate in WD+He star channel may be much lower than the results in this work.}. Future observations 
are needed to provide a larger homogeneous sample, which
would be helpful for understanding the progenitors and explosions of SNe Iax.

\subsection{Parameter study}
\label{sec:parameter}

All the above results are shown with our standard model. However, the predicted 
rates may be sensitive to uncertainties in the binary evolution, such as 
common envelope (CE) evolution or metallicity \footnote{For the effect of more theoretical uncertainties on 
the SN rates and DTDs, see \citet{Clae14}}. The effect
of theoretical uncertainties on the results are addressed in this section. Because 
it is unlikely that SNe Iax are produced from the WD+RG channel, we do not carry out the parameter study 
for this channel here.

\subsubsection{Common envelope efficiency}

The CE-phase plays an essential role in binary evolution in the formation of 
close binaries with compact objects \citep{Ivan13}. Unfortunately, despite 
the enormous efforts of the community, the phenomenon of the CE-phase is
still not understood well \citep{Ivan13}. For reviews of CE evolution, see \citet{Webb08} and \citet{Ivan13}.
After fitting the $\alpha_{\rm{CE}}$-prescription to a population of observed
post-CE binaries, the CE efficiency is estimated to be less than one \citep{Zoro10, De11, Davi12}.
 To investigate
the dependence of the SN rates and DTDs on the uncertainty of the CE phase, we set the combined parameter
$\alpha_{\rm{CE}}\lambda$ to be 0.1, 0.3, 0.5, 1.5 for the WD+MS channel, and set $\alpha_{\rm{CE}}\lambda$ to be 0.5, 1.5
for the WD+He star channel. 

The SN rates and DTDs in the WD+MS channel with different values of $\alpha_{\rm{CE}}\lambda$ are compared 
in Fig.~\ref{Fig:2}.  Lower $\alpha_{\rm{CE}}\lambda$ values tend to 
produce some younger SNe, but the rates in all other cases are lower than for our 
standard case ($\alpha_{\rm{CE}}\lambda$ = 0.5). Here, we point out that the SN rate in the model 
with the lowest $\alpha_{\rm{CE}}\lambda$ value of 0.1 is consistent with the SN Iax rate ($\sim5\%$ of 
the total SN Ia rate) estimated by \citet{Whit14}, and this model 
can produce younger SNe compared to the standard model (see Fig.~\ref{Fig:2}).

\subsubsection{Metallicity}

Figure~\ref{Fig:3} represents the effect of different metallicities ($\rm{Z=}$0.0001, 0.01, 0.02, 0.03) on the  SN rate 
in a single starburst case. More SNe Iax with shorter delay times are produced as the metallicity 
increases. This means the SNe explosion occurs earlier in higher metallicity environments in both the WD+MS and WD+He star 
channels. However, no strong relation between SN rates and metallicity is found.

\subsubsection{Unstable accretion disk}
It has been suggested that the transferred material could form a disk surrounding the 
accreting C/O WD \citep{Osak96, van96}. This accretion disk model has been used to 
explain dwarf–nova outbursts \citep{Osak96}. If the effective temperature 
in the disk falls below the H ionization temperature ($\sim6500\,\rm{K}$), the accretion 
disk may become thermally unstable (\citealt{Osak96, van96}). There 
is a corresponding critical mass-transfer rate of $\dot{M}_{\rm{disk}}$\footnote{We set
 $\dot{M}_{\rm{disk}}=4.3\times 10^{-9}(P_{\rm{orb}}/\rm{4h})^{1.7}\ \rm{M_{\rm{\odot}}yr^{-1}}$ as in \citet{Wang10},
where $P_{\rm{orb}}$ is the orbital period of the binary system.}. The disk is assumed 
to be thermally stable (unstable) if the mass-transfer rate of the donor star, $\dot{M}_{\rm{tr}}$, is 
higher (lower) than $\dot{M}_{\rm{disk}}$. To discuss the effect of the accretion disk, we set 
the accretion rate of the C/O WD to be $\dot{M}_{\rm{2}}=|\dot{M}_{\rm{tr}}|/d$, where $d$ is the 
duty cycle. Here, we simply assume $d=1$ ($=0.01$) if the accretion disk is stable (unstable) (see 
also \citealt{Wang10}). Figure~\ref{Fig:1} shows that more SNe Iax with longer 
delay times are produced if the unstable accretion disk is taken into account, leading to 
increase of the SN rate.

\subsubsection{Mass retention efficiency}

In the SD scenario, only a fairly narrow range in the accretion rate above  $\sim$\,$10^{-7}\,\rm{{M_{\odot}\,yr^{-1}}}$ will
allow stable hydrogen-burning to be attained on the surface of the WD, avoiding a nova explosion. In this work, we adopt 
the prescription of \citet{Hach99} to describe the mass accumulation efficiency of accreting WDs. However some recent 
works concentrated on H/He-burning on accreting WDs obtained different mass retention efficiencies (see, e.g., \citealt{Shen07, 
Wolf13, Pier13, Kato14, Pier14}).

Different mass accumulation efficiencies are usually used in different BPS codes, which may
lead to disagreements about the theoretical predictions in the SD channel (e.g., see
\citealt{Yung00, Han04, Ruit09, Wang10, Menn10, Bour13}).
By studying the influences of three descriptions for the retention efficiencies of WDs on the SN rates and 
DTDs, \citet{Bour13} find that the integrated SN rates can vary by a factor of 3--4 to even more than 
a factor of 100.  \citet{Nele13} have recently collected data from different BPS groups (includes data from our BPS calculations) and 
made a comparison. They find that one of the main differences in the results of the SD models is in
fact from the mass-accretion efficiency of the WD (see \citealt{Nele13}, Nelemans et al., in press). In addition, the 
so-called PopCORN project \citep{Toon14} made a comprehensive comparison of four BPS codes. These authors showed that differences 
between the predictions of BPS codes are not caused by numerical 
effects in the codes, but by different assumptions (such as the mass retention efficiency) in 
stellar or binary evolution \citep{Toon14}. Above all, the predicted SN rates could be different from the results 
presented in this work if different mass-retention efficiencies are 
adopted in BPS our calculations. Therefore, more detailed studies about 
H- and He-burning on accreting WDs are required to place
strong constraints on the uncertainty of retention efficiency.

\section{Interaction between SN ejecta and CSM}
\label{sec:csm}

In the pre-SN accretion phase of the SD model, the donor's stellar winds (symbiotic systems), 
mass loss associated with RLOF (non-conservative mass transfer), or optically thick winds 
from the WD are expected to enrich the CSM around the binary system. After 
the SN explosion, radio, X-ray, and $\rm{H_{\rm{\alpha}}}$ lines are expected to 
be emitted owing to the interaction between the fast-moving SN ejecta and the surrounding 
slow-moving CSM (e.g., \citealt{Chev06}). Therefore, it has long been suggested that radio or X-ray detection of the 
blast wave interaction with the CSM can provide a way to put constraints on the mass-loss 
history of the companion star prior to the SN explosion  and thus on its nature (see, 
e.g., \citealt{Chev06, Imml06, Pana06, Russ12, Chom12, Hore12, Marg12, Marg14, Pere14}).

\begin{figure}
  \begin{center}
    {\includegraphics[width=0.45\textwidth, angle=360]{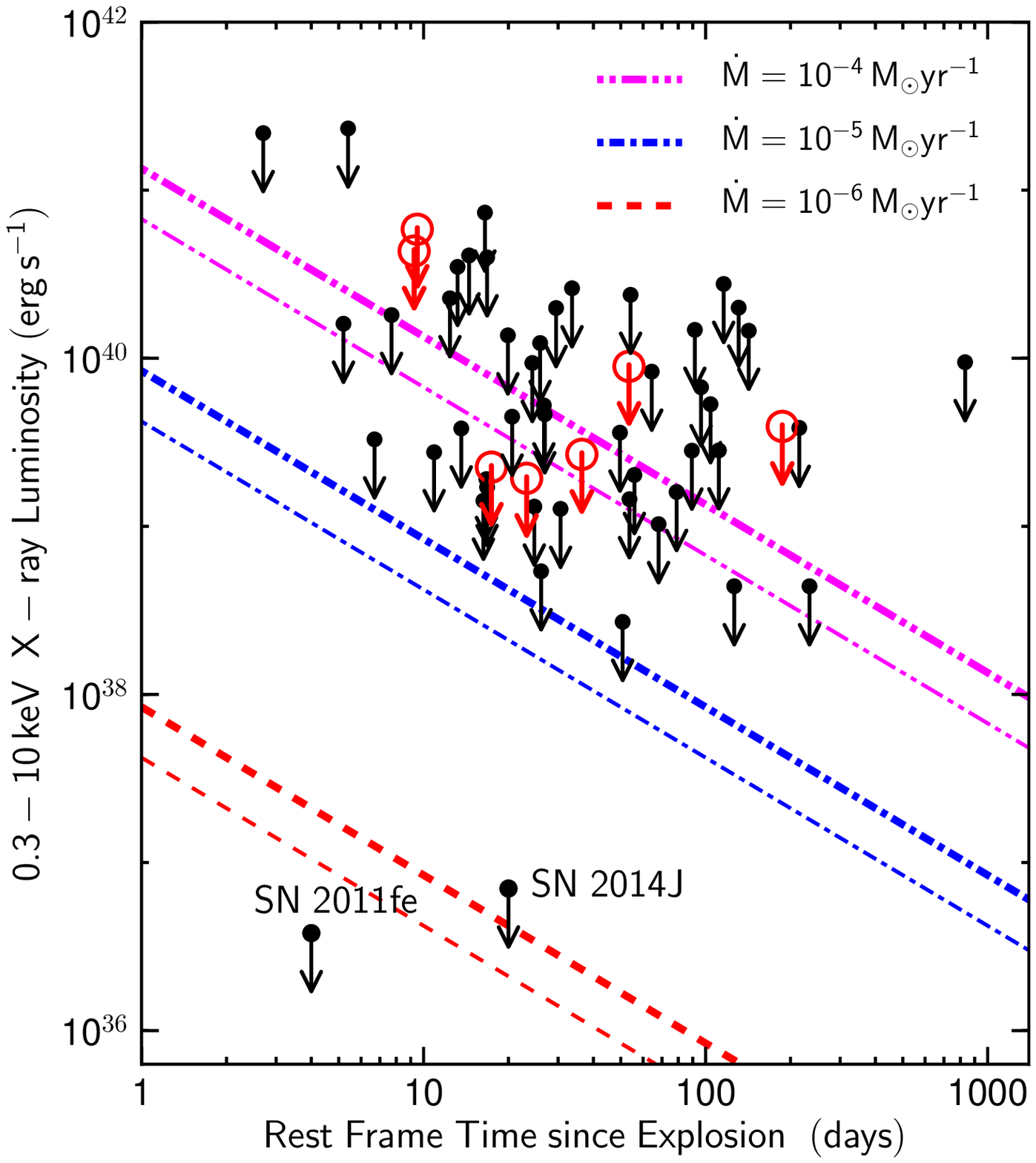}}
  \caption{X-ray observations for SNe Iax (red open circles) and other SNe Ia (black filled circles). Most of the data that was observed 
           by the $\textit{Swift}$ Telescope is taken from \citet{Russ12}. The X-ray emission of SN 2011fe and SN 2014J were observed using the $\textit{Chandra}$ X-ray 
           observations (data obtained from \citealt{Marg12, Marg14}). Theoretical predictions for different
           mass-loss rates $\dot{M}$ (for a wind velocity of $v_{\rm{wind}}=100\,\rm{km\,s^{-1}}$) are shown in different line types. The 
           thin (thick) lines are calculated by assuming a shock velocity of $\rm{10000\,km\,s^{-1}}$ ($\rm{5000\,km\,s^{-1}}$). }
\label{Fig:4}
  \end{center}
\end{figure}

Recently, with a sample of 53 SNe Ia (includes 7 SNe Iax listed in Table~\ref{table:1}) observed by the 
$\textit{Swift}$ X-Ray Telescope (see Fig.~\ref{Fig:4}), \citet{Russ12} calculated a $3\sigma$ upper limit 
for the X-ray luminosity of $1.7\times10^{38}\,\rm{erg\,s^{-1}}$ \citep{Russ12}, which corresponds to an upper limit on the 
mass-loss rate of $1.1\times10^{-6}\,\rm{M_{\odot}\,yr^{-1}}\times\upsilon_{\rm{wind}}/(10\,\rm{km\,s^{-1}})$, where
$\upsilon_{\rm{wind}}$ is the wind velocity. 
However, they only made constraints for normal SNe Ia. Here, again, using the early X-ray 
($0.2$--$10\,\rm{keV}$) observations for SNe Iax reported by \citet{Russ12}, we attempt to examine the validity of the 
SD Ch-mass channel for SN Iax progenitors.

\begin{figure}
  \begin{center}
    {\includegraphics[width=0.45\textwidth, angle=360]{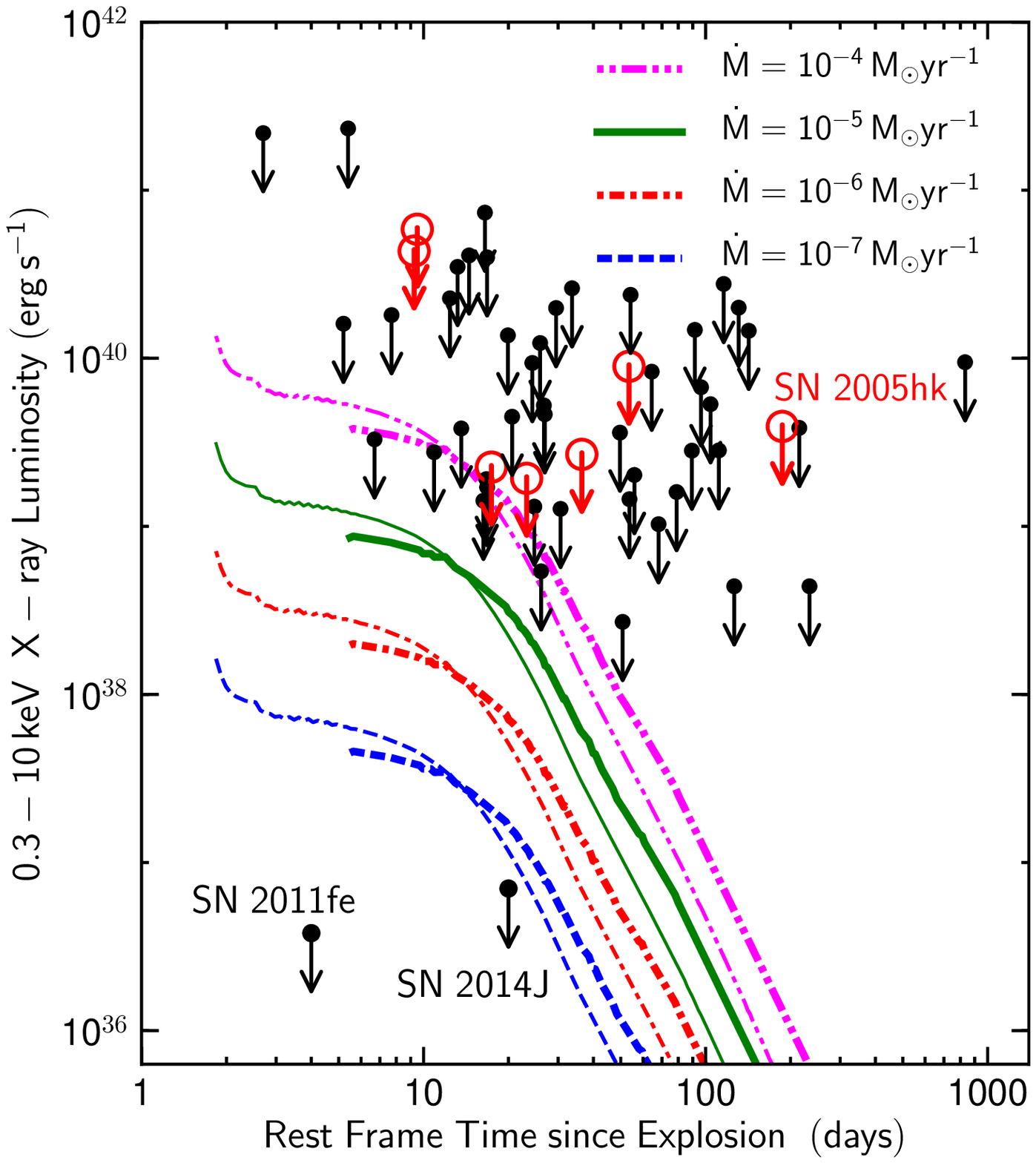}}
  \caption{Similar to Fig.~\ref{Fig:4}, but for the IC X-ray luminosity expected in the case of wind-like CSM ($L_{\rm{IC}}\propto L_{\rm{bol}}$) 
           (see Section~\ref{sec:3.2}).
           Different line types show the results for different mass-loss rates $\dot{M}$ (for a wind velocity 
           of $v_{\rm{wind}}=100\,\rm{km\,s^{-1}}$). The 
           thick lines are calculated by using the observed bolometric luminosity light curve ($L_{\rm{bol}}$) of SN 2005hk \citep{Phil07}. 
           The corresponding results (thin lines) from the `N5def model' of \citet{Krom13} are also shown for a comparison.}
\label{Fig:5}
  \end{center}
\end{figure}

\subsection{Thermal bremsstrahlung model}

In the ejecta-CSM interaction model of \citet{Imml06}, X-ray emissions are ascribed to 
thermal bremsstrahlung. In this model, when assuming material is lost from the binary system 
prior to the SN with a constant mass-loss rate $\dot{M}$ and a wind velocity of $\upsilon_{\rm{wind}}$, the 
thermal X-ray luminosity of the forward shock region is $L=1/(\pi m^{2})\Lambda(T)(\dot{M}/\upsilon_{\rm{wind}})^2(\upsilon_{\rm{s}}t)^{-1}$, 
where $t$ is the time after outburst (see Table~\ref{table:1}), $m$ is the mean mass per particle 
which we take to be $1.8\times10^{-27}\,\rm{kg}$ corresponding to a H+He plasma with solar composition, and $\upsilon_{\rm{s}}$ is the forward 
shock velocity. Here, $\Lambda(T)$ is the cooling function of the heated plasma ($\sim$\,$3\times10^{-23}\,\rm{erg\,cm^{3}\,s^{-1}}$, 
which is assumed by \citealt{Fran96, Imml06}). The profile of the outer ejecta is described by $\rho_{\rm{SN}}\propto r^{\rm{-n}}$, where n 
ranges from 7 to 10 \citep{Kase10}. Moreover, a constant forward shock velocity and $L_{\rm{reverse}}\approx 30\,L_{\rm{forward}}$ was 
assumed in the model of \citet{Imml06}. Here, we assume n=10, as found for SNe arising from compact progenitors
(see \citealt{Matz99, Kase10}). The shock velocity is set to be $\upsilon_{\rm{s}}$ = ($5000, 10\,000\,\rm{km\,s^{-1}}$), the observed 
maximum-light velocity of $2000\lesssim|\upsilon|\lesssim8000\,\rm{km\,s^{-1}}$ in SNe 
Iax \citep{Fole13} provides a lower limit to the shock velocity of the interaction. We adopt a wind 
velocity of $\upsilon_{\rm{wind}} = 100\,\rm{km\,s^{-1}}$. Then, the mass-loss rate of $\dot{M}$ is obtained 
based on X-ray observations of SNe Iax. Our results are shown in Table~\ref{table:1}. 
A comparison between 
the theoretical prediction for thermal bremsstrahlung X-ray emissions and the observations are presented 
in Fig.~\ref{Fig:4}, and a detailed 
discussion is given in Section~\ref{sec:3.4}.

\subsection{Inverse Compton model}
\label{sec:3.2}

It has been also suggested that the X-ray emission from SNe is dominated 
by inverse Compton (IC) scattering of photospheric optical photons by relativistic electrons accelerated by 
the SN shock on a timescale of weeks to a month after the explosion (see 
\citealt{Chev06}). The X-ray luminosity strongly depends on 
the structure of the SN ejecta ($\rho_{\rm{SN}}\propto r^{\rm{-n}}$), the density structure of the CSM and the relativistic electron distribution
responsible for the up-scattering \citep{Chev06, Marg12}. Assuming a wind-like CSM ($\rho_{\rm{CSM}}\propto r^{\rm{-2}}$)
and an ISM-like CSM ($\rho_{\rm{CSM}}=\rm{constant}$), \citet{Marg12} developed two generalized formalisms
to calculate the IC luminosity. Their formalisms are strongly sensitive to the SN bolometric luminosity, $L_{\rm{bol}}$ (see Appendix A of \citealt{Marg12}).
Here, it is assumed electrons are accelerated according to a power-law distribution $n(\gamma)\propto \gamma^{-p}$ with index $p=3$, as 
suggested by the radio observations of SN shocks in Type Ib/Ic explosions (see \citealt{Sode06, Marg14}).
Then, an IC luminosity in the wind-like CSM case is (see \citealt{Marg12}, their Appendix A) 
 \begin{equation}
\begin{split}
    \label{eq:1}
      \frac{\mathrm{d} L_{\rm{IC}}}{\mathrm{d} \nu} \approx 2.1\times10^{-2}\left(\frac{\epsilon_{e}}{0.1}\right)^{-2}\left(\frac{M_{\rm{ej}}}{1.4\,M_{\odot}}\right)^{-0.93}\left(\frac{A}{\rm{g\,cm^{-1}}}\right)^{0.64}
      \\\times\left(\frac{E}{10^{51}\,\rm{erg}}\right)^{1.29}\left(\frac{t}{\rm{s}}\right)^{-1.36}\left(\frac{L_{\rm{bol}}}{\rm{erg\,s^{-1}}}\right)\,\nu^{-1},
\end{split}
   \end{equation}
where $A=\dot{M}/(4\pi\upsilon_{\rm{wind}})$, and $\epsilon_{e}$ is the fraction of thermal energy in the shock used for
the electron acceleration ($\sim0.1$ see \citealt{Chev06}). Theoretical 
IC X-ray luminosities with different mass-loss rates (for a wind velocity of $100\,\rm{km\,s^{-1}}$) are compared to the 
observations in Fig.~\ref{Fig:5}. Here, both the observed bolometric light curve 
of SN 2005hk and the predicted bolometric light curve for the `N5def model' (see Fig.~8 of \citealt{Krom13}) are 
used as input for $L_{\rm{bol}}$. The `N5def model' was shown to provide a good fit to the observations of the SN~2005hk 
event \citep{Krom13}. Moreover, $E=1.34\times10^{50}\,\rm{erg}$, and $M_{\rm{ej}}=0.37\,M_{\odot}$ (these parameters are 
consistent with the N5def model) are used in Equation~\ref{eq:1} to 
estimate the X-ray luminosities. Here, we note that SNe Iax have a 
range of luminosities, ejecta velocities, and inferred ejecta masses \citep{Fole13}. Different bolometric 
luminosity light curves and ejecta masses of different SNe Iax may change the results 
shown in Fig.~\ref{Fig:5}.

\begin{table}
\begin{center}
\caption{X-ray observation for SNe Iax.}\label{table:1}
\begin{tabular}{lcclll}
\hline\hline
     & $t$  & Distance  & $L_{\rm{X}}$  & \multicolumn{2}{c}{$\dot{M}$} \\
Name & $\rm{[days]}$ & $\rm{[Mpc]}$ &  $\rm{[10^{38}erg/s]}$ & \multicolumn{2}{c}{$\rm{[10^{-5}M_{\odot}/yr]}$}\\
     & (1) &   & (2) & (3)& (4) \\ 
\hline
2005hk & 187.0 & 56    & <\,39.3  &   <\,13.1&  <\,9.3 \\
2008A  & 53.5  & 70    & <\,89.5  &   <\,10.7&  <\,7.5 \\
2008ae & 9.26  & 127   & <\,434   &   <\,9.8&  <\,6.9 \\
2008ge & 23.2  & 16    & <\,19.3  &   <\,3.3&  <\,2.3 \\
2008ha & 36.4  & 22.5  & <\,26.8  &   <\,4.8&  <\,3.4 \\
2010ae & 17.4  & 17    & <\,22.5  &   <\,3.1&  <\,2.2 \\
2010el & 9.5   & 30.19 & <\,585.4 &   <\,11.5&  <\,8.1 \\
\hline
\end{tabular}
\tablefoot{SNe Iax were observed by the $\textit{Swift}$ X-Ray Telescope (XRT). The data used here 
           obtained from \citet{Russ12}. (1) $t$ is the time after outburst (see \citealt{Russ12});
           (2) $3\sigma$ upper limit to the 0.2--10 KeV X-ray band luminosity; (3) upper-limit mass-loss rate of 
           the progenitor system (for wind velocity of $\upsilon_{\rm{wind}} = 100\,\rm{km\,s^{-1}}$), for which a power
           law index of the SN ejecta of n= 10 and $\upsilon_{\rm{s}} = 10\,000\,\rm{km\,s^{-1}}$ are adopted; (4) similar 
           to (3), but for $\upsilon_{\rm{s}} = 5000 \,\rm{km\,s^{-1}}$ case. \\
          }

\end{center}
\end{table}

\begin{figure}
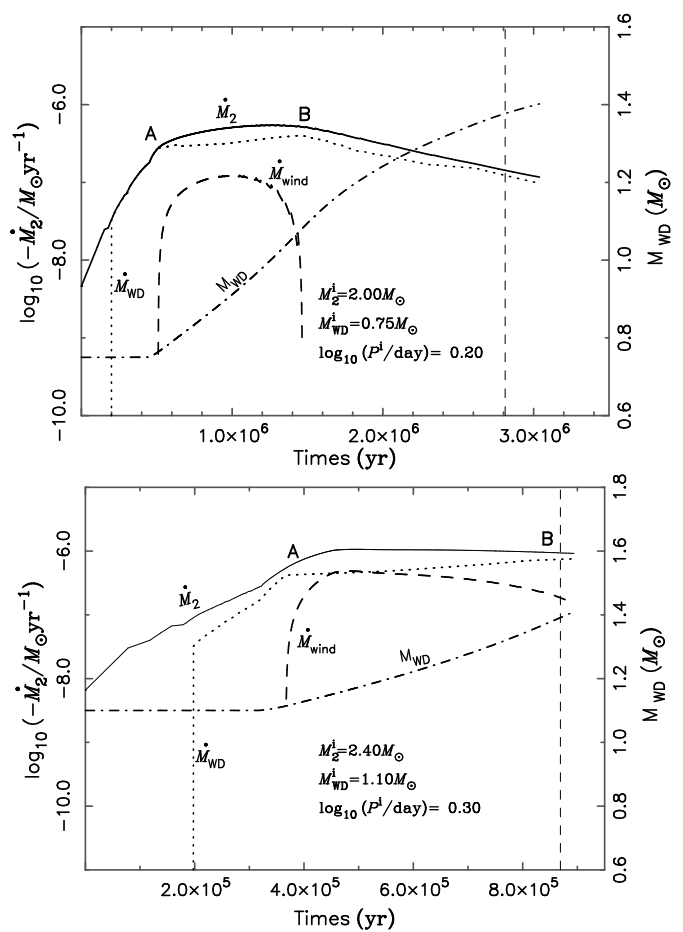

  \begin{center}
    {\includegraphics[width=0.33\textwidth, angle=270]{f6a.eps}}
    \vspace{0.2in}
   {\includegraphics[width=0.33\textwidth, angle=270]{f6b.eps}}
  \caption{Evolution of the mass-accretion rate of the WD ($\dot{M}_{\rm{2}}$, solid curve), mass-loss rate in the optically thick wind phase
           ($\dot{M}_{\rm{wind}}$, dashed curve, i.e., the phase from `$\rm{A}$' to `$\rm{B}$') and mass growth rate of the  WD ($\dot{M}_{\rm{WD}}$, dotted
           curve) as a function of time for the WD+MS channel (left-hand axis). The dot-dashed 
           curve shows the evolution  of the WD mass (right-hand axis). The vertical dashed curve again indicates the time of the explosion.
           The top panel gives an example where there is a delay between the production of the wind and
           the SN explosion, while the bottom panel corresponds to a model that is in the wind phase at the time of explosion.
}
\label{Fig:6}
  \end{center}
\end{figure}

\begin{figure}
  \begin{center}
    {\includegraphics[width=0.45\textwidth, angle=360]{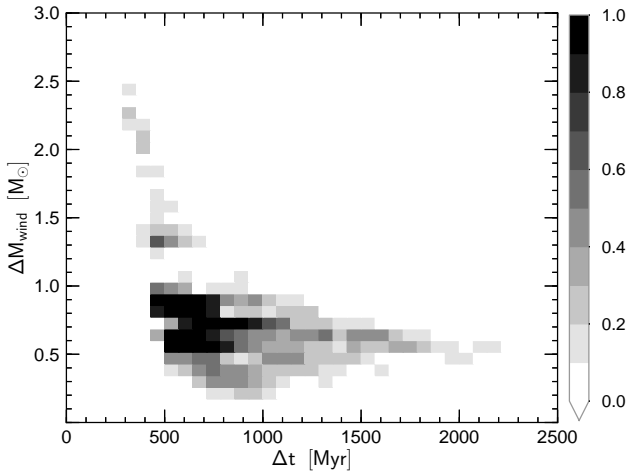}}
  \caption{Distributions of all the WD+MS models in the plane of
           ($\Delta \rm{t}$, $\Delta \rm{M_{wind}}$), where $\Delta \rm{M_{wind}}$ is the mass lost during 
            the optically thick stellar wind phase of the WD binary, and $\Delta \rm{t}$ is the SN delay time.}
\label{Fig:7}
  \end{center}
\end{figure}

\subsection{Theoretical pre-SN mass loss}

\label{sec:3.3}

In our calculations, an accreting C/O WD is assumed to blow an optically thick wind if 
$\dot{M}_{\rm{2}}>\dot{M}_{\rm{cr}}$. This leads to a typical mass-loss rate that is less than $\rm{10^{-6}\,M_{\odot}yr^{-1}}$, 
but may be up to $\rm{10^{-4}\,M_{\odot}yr^{-1}}$ in extreme and rare cases (see also \citealt{Han06}). 
The wind velocity is expected to be several $10^{3}\,\rm{km\,s^{-1}}$ 
(or a few $10^{2}\,\rm{km\,s^{-1}}$ ) if it comes from the neighbourhood of the WD (or comes from the circumbinary 
envelope).  Figure~\ref{Fig:6} presents two typical examples of mass-loss history in our detailed binary evolution 
calculations for the WD+MS models. It shows the unprocessed material lost from the binary system in an 
optically thick wind with a mass-loss rate of $\dot{M}_{\rm{wind}}=|\dot{M}_{\rm{2}}|-\dot{M}_{\rm{cr}}$ during the phase 
from `A' to `B'. The amount of mass lost during the optically thick wind phase as a function of the delay time is 
also displayed in Fig.~\ref{Fig:7}. A large amount of mass loss from the system in the pre-SN phase can produce 
the inferred H-rich circumstellar environment.

\subsection{Observational constraints on mass-loss rates}
\label{sec:3.4}

Figures~\ref{Fig:4} and \ref{Fig:5} show that theoretical predictions (see also Table~\ref{table:1}) for the case of 
$\dot{M}/v_{\rm{wind}}\sim10^{-4}\,\rm{M_{\odot}\,yr^{-1}/(100\,km\,s^{-1})}$ roughly match the 
upper-limit X-ray luminosities from the observations of SNe Iax. However, one should keep in
mind that the observed X-ray luminosities of SNe Iax only provide an upper limit 
on the pre-SN mass-loss rate of SN Iax progenitors. As mentioned above, although the extreme and rare 
case with a high mass-loss rate of $\rm{10^{-4}\,M_{\odot}yr^{-1}}$ can occur in our 
calculations (see also \citealt{Han06}), mass-loss rates of most SNe Iax progenitor systems
in our models are less than $\rm{10^{-6}\,M_{\odot}yr^{-1}}$. Such 
mass-loss rates are much lower than the upper-limit
ones obtained from the X-ray observations, thus suggesting that the SD H/He-accreting
Ch-mass model cannot be excluded from the possible progenitors of SNe Iax by current X-ray 
observations.

Furthermore, only a small fraction of all models (for example, less than 10\% of models in WD+MS channel) is 
still in the optically think wind phase at the time of the SN explosion in our BPS calculations. Most of our models have a delay 
between the production of the wind and the SN ($\sim10^5$--$10^6\,\rm{yr}$ in WD+MS channel, see 
Fig.~\ref{Fig:6}). This delay could produce a low-density CSM around the system when the SN explodes, 
thus leading to a non-detection (weak) of the X-ray emission at early times after the SN explosion. However, this 
case would create a large cavity surrounding the WD at the SN explosion and may leave a clear imprint on 
supernova remnants (SNRs, see \citealt{Bade05, Bade07}).

Taking all this into consideration, we can safely suggest that the upper-limit constraints on the pre-SN 
mass-loss rates obtained from current early X-ray observations for SNe Iax do not exclude the 
studied SD Ch-mass channels as progenitors of SNe Iax. However, we note that the theoretical X-ray luminosities 
shown in Figs.~\ref{Fig:4} and \ref{Fig:5} are obtained from analytic methods with some simple assumptions. 
To obtain strict constraints on the 
mass-loss rates of the progenitor models, detailed numerical hydrodynamical simulations for the 
interaction between the blast wave of SN and the CSM (e.g., \citealt{Dimi14}) are required, coupled with a suitable 
explosion model for SNe Iax (such as those of the failed deflagration model in \citealt{Jord12, Krom13}) and the detailed pre-SN 
mass-loss history (see Fig.~\ref{Fig:6}).

In the binary evolution of the SD Ch-mass scenarios, it was also suggested that a small fraction of the transferred mass ($\lesssim1\%$) may 
be lost at the outer Lagrangian points of the binary system with a wind velocity of several $100\,\rm{km\,s^{-1}}$ (up 
to $\sim600\,\rm{km\,s^{-1}}$, see \citealt{Deuf99}) when the WD undergoes steady nuclear burning (see also \citealt{Chom12, Marg14}). 
The expected mass-loss rates from this case are still below the upper-limit mass-loss constraints from X-ray
observations for SNe Iax. This model thus also cannot be ruled out as a possible scenario for producing the progenitors of SNe Iax. Moreover, the accreting WD in a binary system 
can experience recurrent novae before it increases its mass to the Ch-mass to cause the SN explosion. \citet{Wood06} suggested 
that recurrent novae produce a low-density cavity in the immediate CSM, which may lead to 
a weak X-ray emission after the SN. It is also suggest that radio emission may be generated as 
the SN shockwave crashes through a previously ejected shell \citep{Wood06}.

\section{Discussion}

\label{sec:discussion}

\subsection{He lines}

Two SNe Iax (SN 2004cs and 2007J) were found to show evidence for He in their
progenitor system \citep{Fole13}. In present work, we suggest that some
SNe Iax are likely to occur in progenitor systems that consist of a He-star companion 
and a C/O WD (see also \citealt{Fole13}). This scenario may explain He lines 
seen in SN Iax spectra.

Current failed deflagration models of SNe Iax are based on C/O Ch-mass WD \citep{Jord12, Krom13, Fink13}. 
The weak deflagration explosion of a C/O WD that contains a thick He shell\footnote{This He shell 
may form following both He accretion from a He-star companion
or H accretion and burning into helium on a WD.} can probably explain 
the He lines seen in SN Iax spectra. We encourage numerical simulations for the weak deflagration explosions of 
Ch-mass WDs to explore this He-shell model further.

\subsection{Detection of stripped H or He}

In the SD Ch-mass model, the mass donors are H- or He-rich stars. After the SN explosion, the SN ejecta 
hits the companion and removes the material from its outer layers through the ablation (SN heating) and 
the stripping (momentum transfer) mechanisms \citep{Liu12, Pan12}. 

Recent three-dimensional impact simulations for SN Iax  showed that only a small amount ($\lesssim 0.01\,\rm{M_{\odot}}$) 
of H-rich material is removed from the MS companion by the SN impact \citep{Liu13b}.  If the donor star is a He 
star, a lower He mass is expected to be removed by the SN impact (see \citealt{Pan10, Pan12, Liu13c}) because 
the He star is much more compact than the MS companion. This indicates 
that the H or He lines are likely to be hidden in the late-time spectra of SNe Iax. 
A higher explosion energy would lead to a higher stripped H or He 
mass. Thus, more luminous SNe Iax seem more likely to show stripped H or He.
However, whether or not H or He lines from matter stripped from the companion star 
is detectable can only be answered by performing sophisticated radiative-transfer 
simulations on the abundance structure of explosion models.

\subsection{Post-explosion survivor}

A weak deflagration of a C/O WD does not burn the entire WD, but leaves 
behind a $\sim1.0\,\rm{M_{\odot}}$ bound remnant \citep{Jord12, Krom13, Fink13}. 
Unfortunately, this bound remnant cannot be spatially resolved until late times 
in hydrodynamical simulations due to the strong expansion of the SN ejecta. Therefore, whether  
the binary system could be destroyed after the SN explosion is still unclear \citep{Liu13b}. If 
the bound remnant of the Ch-mass WD receives a high kick velocity 
from the explosion, a surviving WD and a companion star with a peculiar spatial velocity and peculiar abundances
can be indicators of this studied progenitor scenario \citep{Jord12, Krom13, Fole14}. Otherwise, the binary system would survive 
the SN explosion \citep{Liu13b}. Further investigations are still required to 
strongly constrain the kick velocity of the bound remnant. However, fully resolving the detailed structure
of the bound remnants is a prerequisite for this investigation \citep{Fink13}.

\subsection{X-ray emission from accreting WDs}

If SNe Iax come from the SD Ch-mass scenario, the accreting 
WDs are expected to undergo a supersoft X-ray source (SSS) phase before the SN explosions when their masses are 
increasing \citep{Van92, Iben94, Kaha97, Yoon03, Nomo07}. Comparing the observed SSSs in
galaxies of different morphological types with expectations from the SD Ch-mass
model can provide constraints on the progenitors of SNe Iax (e.g., \citealt{Di10a, Di10b, Gilf10}).

There are still large uncertainties, however, on the theoretical X-ray luminosity of the SSSs, such as
the atmospheric models of accreting WDs and absorption of 
soft X-rays \citep{Hach10}. Depending on the different mass-accretion rates, 
the accreting WDs would undergo novae, wind evolution, or SSS phases. 
It is shown that the SSS phase is only a short time, because most of 
the accreting WDs in the SD scenario spend a large portion of time in the optically 
thick wind phase and the recurrent nova phase \citep{Hach10}.

\subsection{Other possible models}

It has recently been suggested that a WD that accretes material from a He-rich companion star 
is more likely to produce a double-detonation SNe Ia (i.e., double-detonation scenario [DDS],  
see \citealt{Fink07, Fink10, Woos11b, Sim12, Moll13}). 
It was also suggested that SNe Iax may come from the DDS
since the predicted rates and DTDs from the DDS (which are also shown in Figs.~\ref{Fig:1} 
and \ref{Fig:2}) are comparable with the observations \citep{Wang13}.
However, current simulations \citep{Fink10, Krom10, Sim12, Woos11b} for the DDS show that this explosion  
struggles to reproduce the characteristic features of SNe Iax. For example, the low ejecta velocity 
of SNe Iax, strong mixing of their explosion ejecta, and the full diversity of SNe Iax are hard
to reproduce. However, the details of the DDS 
still require future developments and numerous complications remain to be solved in such a model.

Additionally, a so-called ``.Ia'' model was proposed to explain faint rapidly rising thermonuclear 
SNe \citep{Bild07, Shen10}. In this model, He is accreted onto the surface of a WD, leading to a single 
He layer detonation to cause an explosion of the WD. With this model, some observables of SNe Iax can be 
roughly matched such as the faint, fast-rising and declining features in SN 2008ha. However, this ``.Ia''  
model primarily yields heavy $\alpha$-chain elements and unburnt He, but it is difficult to produce 
intermediate-mass elements \citep{Shen10}. Moreover, this model fails to  
reproduce the full diversity of SNe Iax (see \citealt{Fole09}).

\subsection{SN 2008ha-like events}

Observationally, a very low $^{56}\rm{Ni}$ mass of $\sim0.003\,\rm{M_{\odot}}$ was derived 
for two SNe Iax, SN 2008ha and SN 2010ae \citep{Fole09, Stri14}. This low $^{56}\rm{Ni}$ mass cannot be explained easily by 
current published models. \citet{Fole09, Fole10} suggest that SN 2008ha 
may be produced from the deflagration of a sub-Chandrasekhar-mass WD, and it was also proposed as 
the result of the core-collapse of a massive star \citep{Vale09, Mori10}. A minimum 
$^{56}\rm{Ni}$ mass from current simulations for the failed deflagration explosions is $\sim0.035\,\rm{M_{\odot}}$
\citep{Fink13}, which is more than a factor of 10 greater than the observed value for 
SN 2008ha-like events. Very recently, \citet{Fole14} have detected a source that coincident with the position 
of SN 2008ha with $M_{\rm{F814W}}=-5.4\,\rm{mag}$. They suggest that this source may be 
the stellar donor or the remnant predicted by the failed deflagration 
model \citep{Jord12, Krom13, Fink13}. However, this still needs future observations for confirmation.

\section{Summary and conclusions}
\label{sec:summary}

Assuming weak deflagration explosions of C/O Ch-mass WDs in the H/He-accreting SD scenario can 
produce SNe Iax, we have compared predicted SN rates and delay times from BPS calculations to 
the observations of SNe Iax. Moreover, assuming the SNe Iax are a homogeneous class of objects, 
we calculated upper limits on the mass-loss rates from the progenitor systems. We tried to 
examine constraints on the H/He-accreting SD models as progenitors of SNe Iax. The main results 
are summarized as follows.

\begin{itemize}

\item[1.]  The long delay times of $\gtrsim3\,\rm{Gyr}$ and low SN rates of $\sim$$3\times10^{-5}\,\rm{yr^{-1}}$ from 
           the WD+RG channel indicate this scenario is unlikely to produce SNe Iax. 

\item[2.]  The SNe Iax from the WD+He~star channel contribute about $\sim$\,$9\%$ of the total SN Ia
           rate. The DTDs in the WD+He star channel ($<100\,\rm{Myr}$) allow for
           the observed SNe Iax to favour late-type galaxies.

\item[3.]  The SN rate ($\sim$\,$43\%$ of the total SN Ia rate) from the WD+MS channel is comparable 
           to the estimated rate ($30_{-13}^{+17}\%$ of the total SNe Ia rate) of \citet{Fole13}, but it is much 
           higher than the one ($5.6_{-3.7}^{+17}\%$ of the total SNe Ia rate) supported by \citet{Whit14}. 

\item[4.]  Current 
           observational constraints on the ages of progenitor systems ($<80\,\rm{Myr}$, see \citealt{Lyma13, Fole14}) are 
           less favourable for the delay times in the WD+MS channel ($\sim250\,\rm{Myr}$--$1\,\rm{Gyr}$). 
           However, at least one SN Iax event (SN 2008ge) is hosted by a S0 galaxy with no signs of 
           star formation.
\item[5.]  The upper limits for the mass-loss rates of progenitor systems derived from X-ray 
           observations for SNe Iax are $\dot{M}\lesssim \rm{a\ few}\times10^{-4}\,\rm{M_{\odot}\,yr^{-1}}$ (for
           wind velocity $v_{w}=100\,\rm{km\,s^{-1}}$). Considering that the typical pre-SN mass-loss rates 
           of H- or He-accreting SD models are $\sim$$10^{-6}\,\rm{M_{\odot}yr^{-1}}$, we conclude that 
           current X-ray observations cannot rule out that the SD Ch-mass model offers possible progenitor scenarios for SNe Iax.
\item[6.]  We suggest that some SNe Iax may come from weak deflagration explosions of C/O WDs in the SD Ch-mass scenario,
           especially in the WD+He star channel. However, this SD Ch-mass channel is still too difficult 
           to be the most common progenitor scenario for SNe Iax. 
\end{itemize}

The problem of the progenitors of SNe Iax is still poorly constrained. No single published 
model is able to explain all the observational features and full diversity of SNe Iax. If SNe 
Iax are assumed to be generated from the same origin, a new progenitor paradigm 
may be needed. Otherwise, a combination of current models might be an alternative scenario for SNe Iax. 
Future observations providing a bigger sample of SNe Iax will be very helpful for constraining the rates of these events and their host-galaxy 
morphology distributions.

\section*{Acknowledgments}

      We acknowledge the referee Carles Badenes for his valuable 
      comments and suggestions that helped us to improve the paper. We thank Yan-Rong Gong for her
      helpful suggestions. We thank Markus Kromer for providing bolometric-luminosity data of SN 2005hk and their
      N5def model. This work is supported by the Alexander von Humboldt Foundation. 
      RJS is the recipient of a Sofja Kovalevskaja Award from the Alexander von Humboldt Foundation.
      TJM is supported by Japan Society for the Promotion of Science
Postdoctoral Fellowships for Research Abroad (26\textperiodcentered 51).
      BW acknowledges support from the 973 programme of China (No. 2014CB845700), the NSFC
      (Nos. 11322327, 11103072, 11033008 and 11390374), the Foundation 
      of Yunnan province (No. 2013FB083).


\begin{appendix}
\section{Accretion efficiencies of WDs}
\label{sec:appendix1}

In our detailed binary evolution calculations, an optically thick-wind assumption 
of \citet{Hach96, Hach99} is used to describe the mass growth of a C/O WD by accretion 
of H-rich material from the donor star, and the prescription of \citet{Kato04} is 
inserted into the code for the mass accumulation efficiency of the He-shell flashes 
on to the WDs.

\subsection{The H-rich donor channel}

For H-rich companion star (MS, sub-giant, and RG), the mass growth rate of the C/O 
WD is set to be $\dot{M}_{\rm{WD}}=\eta_{\rm{He}} \dot{M}_{\rm{He}} = \eta_{\rm{He}} \eta_{\rm{H}} \dot{M}_{\rm{2}}$, 
where  $\dot{M}_{\rm{He}}$ is the mass-growth rate of the He layer under the H-shell
burning.  $\eta_{\rm{H}}$ is the mass-accumulation efficiency for hydrogen
burning and is controlled by

 \begin{equation}
    \label{eq:a1}
\eta_{\rm{H}} = \left\{ \begin{array}{ll}
\dot{M}_{\rm{cr,H}}/|\dot{M}_{\rm{2}}|,  & |\dot{M}_{\rm{2}}| > \dot{M}_{\rm{cr,H}},  \\
1, & \dot{M}_{\rm{cr,H}} \geqslant \dot{M}_{\rm{2}} \geqslant\frac{1}{8}\dot{M}_{\rm{cr,H}},\\
0, & \dot{M}_{\rm{2}} < \frac{1}{8}\dot{M}_{\rm{cr,H}},
\end{array} \right.
  \end{equation}
where $\dot{M}_{\rm{cr,H}} = 5.0\times10^{-7}(1.7/X-1)(M_{\rm{WD}}/\rm{M_{\odot}}-0.4)\,\rm{M_{\odot}yr^{-1}}$ is 
the critical accretion rate for stable hydrogen
burning, $X$ is the H mass fraction, $M_{\rm{WD}}$ the mass of the accreting WD, 
and $\dot{M}_{\rm{2}}$ the mass-accretion rate of the WD. Also, ${\eta}_{\rm{He}}$  is the mass-accumulation efficiency for 
He-shell flashes, and its value is taken from \citet{Kato04}. 

The optically thick wind \citep{Hach96} is assumed to blow off all unprocessed material if $|\dot{M}_{\rm{2}}|$ is 
greater than $\dot{M}_{\rm{cr,H}}$, and the lost material is assumed to take 
away the specific orbital angular momentum of the accreting WD (the mass loss in the donor’s wind is supposed to be negligible, but its
effect on the change in the orbital angular momentum, i.e. magnetic braking is included).

\subsection{The He-rich donor channel}

For He donor star model, the mass growth rate of the C/O WD $\dot{M}_{\rm{WD}}=\eta_{\rm{He}} \dot{M}_{\rm{2}}$. 
The $\eta_{\rm{He}}$ is the mass-accumulation efficiency for He burning, which 
is set to be 
 
 \begin{equation}
    \label{eq:a3}
\eta_{\rm{He}} = \left\{ \begin{array}{ll}
\dot{M}_{\rm{cr,He}}/|\dot{M}_{\rm{2}}|,  & |\dot{M}_{\rm{2}}| > \dot{M}_{\rm{cr,He}},  \\
1, & \dot{M}_{\rm{cr,H}} \geqslant \dot{M}_{\rm{2}} \geqslant\dot{M}_{\rm{st}},\\
\eta_{\rm{He}}', & \dot{M}_{\rm{st}} \geqslant \dot{M}_{\rm{2}} \geqslant\frac{1}{8}\dot{M}_{\rm{low}},\\
0, & \dot{M}_{\rm{2}} < \dot{M}_{\rm{low}},
\end{array} \right.
  \end{equation}
where $\dot{M}_{\rm{cr,He}} = 7.2\times10^{-6}(M_{\rm{WD}}/\rm{M_{\odot}}-0.6)\,\rm{M_{\odot}yr^{-1}}$ is 
the critical accretion rate for stable He burning; $\dot{M}_{\rm{st}}$ is the minimum accretion
rate for stable He-shell burning \citet{Kato04}; $\dot{M}_{\rm{low}}=4\times10^{-8}\,\rm{M_{\odot}yr^{-1}}$ 
is the minimum accretion rate for weak He-shell flashes \citep{Woos86}; ${\eta}_{\rm{He}}'$ obtained from 
linearly interpolated from a grid computed by \citet{Kato04}.

\section{Evolutionary channels of progenitors}
\label{sec:appendix2}
\subsection{The WD+MS and WD+RG channels}

According to evolutionary phase of the primordial primary at the beginning of the first RLOF, there are three
channels that can form WD+MS systems, and one channel that can form WD+RG systems.

(1) The primordial primary first fills its Roche lobe when it is in the Hertzsprung gap (HG) or first giant 
    branch (FGB) stage, and the binary system goes through the CE-phase. After the CE ejection, the primary 
    becomes a He star to form a He star+MS system. The He star continues to evolve and may fill its Roche 
    lobe at its RG stage and transfer its remaining He-rich envelope onto the surface of the MS companion 
    star, eventually leading to the formation of a C/O WD+MS system. For this channel, the initial masses of 
    the primary and the secondary at the zero age main-sequence (ZAMS) are $M_{\rm{1,i}}\sim4.0$--$7.0\,\rm{M_{\odot}}$ and 
    $M_{\rm{2,i}}\sim1.0$--$2.0\,\rm{M_{\odot}}$, the initial orbital period of the binary system is $P^{\rm{i}}\sim5$--$30\,\rm{days}$.

(2) If the primordial primary fills its Roche lobe at the early asymptotic giant branch (EAGB) and the system 
    goes through the CE phase, a close He RG+MS binary may then be produced after CE is ejected. The He RG 
    further fills its Roch lobe and a C/O WD+MS system can be formed after the RLOF. For this scenario, 
    $M_{\rm{1,i}}\sim2.5$--$6.5\,\rm{M_{\odot}}$, $M_{\rm{2,i}}\sim1.5$--$3.0\,\rm{M_{\odot}}$, and $P^{\rm{i}}\sim200$--$900\,\rm{days}$. 

(3) The primordial primary fills its Roche lobe at the thermal pulsing asymptotic giant branch (TPAGB) stage. A 
    CE is formed owing to dynamically unstable RLOF. After the CE ejection, a C/O WD+MS system can be produced. 
    The C/O WD+MS system continues to evolve, depending on different initial masses of binary components and 
    initial binary separation, the MS companion can fill its Roche lobe in its MS phase ($M_{\rm{1,i}}\sim4.5$--$6.5\,\rm{M_{\odot}}$, 
    $M_{\rm{2,i}}\sim1.5$--$3.5\,\rm{M_{\odot}}$, $P^{\rm{i}}>1000\,\rm{days}$) or when it
    evolves to the RG phase (i.e., forming the WD+RG system, $M_{\rm{1,i}}\sim5.0$--$6.5\,\rm{M_{\odot}}$, 
    $M_{\rm{2,i}}\sim1.0$--$1.5\,\rm{M_{\odot}}$, $P^{\rm{i}}>1500\,\rm{days}$ ), finally leading to SN explosion.

\subsection{The WD+He star channel}

\begin{table}
\begin{center}
\caption{Initial parameters of the binary system at ZAMS.}\label{table:b1}
\begin{tabular}{lccl}
\hline\hline
         &\multicolumn{3}{c}{WD+He star}  \\
Scenario & $M_{\rm{1,i}}$    & $M_{\rm{1,i}}$    & $P^{\rm{i}}$ \\
         & $[\rm{M_{\odot}}]$ & $[\rm{M_{\odot}}]$  & $[\rm{days}]$ \\
 \hline
(1) & 5.0--8.0 & 2.0--6.5 & 10--40  \\
(2) & 6.0--6.5 & 5.5--6.0 & 300--1000 \\
(3) & 5.5--6.5 & 5.0--6.0 & >1000 \\
\hline
\end{tabular}
\tablefoot{$M_{\rm{1,i}}$ and  $M_{\rm{2,i}}$ are the initial masses of 
           the primary and the secondary at the ZAMS, and
           $P^{\rm{i}}$ is the initial orbital period of the binary system. \\
          }

\end{center}
\end{table}

There are three evolutionary scenarios to form WD+He star systems. The typical phases that the binary
system goes through are shown below. The typical masses of the primary and secondary at the 
ZAMS and initial orbital period of the binary list in Table~\ref{table:b1}.

(1) MS+MS $\overset{}{\underset{}{\text{\scalebox{2}[1]{$\rightarrow$}}}}$  Subgiant/FGB+MS 
      $\overset{\rm{Stable\ RLOF}}{\underset{}{\text{\scalebox{4}[1]{$\rightharpoondown$}}}}$ He~star+MS
 $\overset{}{\underset{}{\text{\scalebox{2}[1]{$\rightarrow$}}}}$ He RG+MS$\overset{\rm{Stable\ RLOF}}{\underset{}{\text{\scalebox{4}[1]{$\rightharpoondown$}}}}$  C/O WD+MS 
 $\overset{}{\underset{}{\text{\scalebox{2}[1]{$\rightarrow$}}}}$ C/O WD+Subgiant/FGB
  $\overset{\rm{Unstable\,RLOF}}{\underset{}{\text{\scalebox{4}[1]{$\rightharpoondown$}}}}$  CE 
  $\overset{\rm{CE\ ejection}}{\underset{}{\text{\scalebox{4}[1]{$\rightharpoondown$}}}}$  C/O WD+He star
 $\overset{}{\underset{}{\text{\scalebox{2}[1]{$\rightarrow$}}}}$ SN event.

(2) MS+MS$\overset{}{\underset{}{\text{\scalebox{2}[1]{$\rightarrow$}}}}$  EAGB+MS 
      $\overset{\rm{Unstable\ RLOF}}{\underset{}{\text{\scalebox{4}[1]{$\rightharpoondown$}}}}$ CE
 $\overset{\rm{CE\ ejection}}{\underset{}{\text{\scalebox{4}[1]{$\rightharpoondown$}}}}$ He RG+MS$\overset{\rm{Stable\ RLOF}}{\underset{}{\text{\scalebox{4}[1]{$\rightharpoondown$}}}}$  C/O WD+MS 
 $\overset{}{\underset{}{\text{\scalebox{2}[1]{$\rightarrow$}}}}$ C/O WD+Subgiant/FGB
  $\overset{\rm{Unstable\,RLOF}}{\underset{}{\text{\scalebox{4}[1]{$\rightharpoondown$}}}}$  CE 
  $\overset{\rm{CE\ ejection}}{\underset{}{\text{\scalebox{4}[1]{$\rightharpoondown$}}}}$  C/O WD+He star
 $\overset{}{\underset{}{\text{\scalebox{2}[1]{$\rightarrow$}}}}$ SN event.

(3) MS+MS$\overset{}{\underset{}{\text{\scalebox{2}[1]{$\rightarrow$}}}}$   TPAGB+He-core burning star 
      $\overset{\rm{Stable\ RLOF}}{\underset{}{\text{\scalebox{4}[1]{$\rightharpoondown$}}}}$  CE 
  $\overset{\rm{CE\ ejection}}{\underset{}{\text{\scalebox{4}[1]{$\rightharpoondown$}}}}$  C/O WD+He star
  $\overset{}{\underset{}{\text{\scalebox{2}[1]{$\rightarrow$}}}}$ SN event.

\end{appendix}

\bibliographystyle{aa}

\bibliography{ref}

\end{document}